Diffraction tomography on curved boundaries: A projection-based approach

*Gregory T. Clement*

Department of Biomedical Engineering, Cleveland Clinic Foundation, 9500 Euclid Ave/ND 20, Cleveland, Ohio 44195: clemeng@ccf.org

## Abstract

An approach to diffraction tomography is investigated for two-dimensional image reconstruction of objects surrounded by an arbitrarily-shaped curve of sources and receivers. Based on the integral theorem of Helmholtz and Kirchhoff, the approach relies upon a valid choice of the Green's functions for selected conditions along the (possibly-irregular) boundary. This allows field projections from the receivers to an arbitrary external location. When performed over all source locations, it will be shown that the field caused by a hypothetical source at this external location is also known along the boundary. This field can then be projected to new external points that may serve as a virtual receiver. Under such a reformation, data may be put in a form suitable for image construction by synthetic aperture methods. Foundations of the approach are shown, followed by a mapping technique optimized for the approach. Examples formed from synthetic data are provided.

## 1. Introduction

Although diffraction tomography has been described in various generalized forms [1–4], reconstructions are typically performed along separable boundaries [5–7], *i.e.* data are acquired over a curve or surface formed by holding constant one or more dimension of a given coordinate system. This condition need not be restrictive, provided such a surface (or line, in 2D) can be accessed. However, for a large class of problems, placement of both transmitters and receivers is limited by reachable geometry and coupling ability. Such limitations are particularly prevalent in acoustics where the boundary of the structure being investigated is often irregularly-shaped. Some examples would include the human body, biologic samples, and many geologic structures.

A method for reconstructing images using an arbitrary configuration of transmitters and receivers was described in pioneering work by Devany and Beylkin [8], who demonstrated such reconstruction can be performed under the key condition that the local radius of curvature is much larger than a wavelength. Gelius also described a general technique valid for curved acquisitions, based on a paraxial approximation [9]. The present work seeks to build on the arbitrary approach by introducing an alternative method that does not require explicit restrictions on curvature. The method is based on the integral theorem of Helmholtz and Kirchhoff [10] which, with care, can be applied to quite general shapes. The current approach applies Dirichlet conditions to the irregular boundary, such that successful implementation is dependent upon the choice and validity of the particular Green's functions selected to satisfy conditions.

Given a sufficiently large internal region, it will be shown that selection of the Green's functions necessary to satisfy boundary conditions can be straightforward. Examples considered here in Section 3 are limited to such a case. However, the formulation itself is not restrictive of size, and for smaller regions, boundary element [11–14], null-field[15,16], or series solutions [17] could be used.

The scope of the present work is further limited to the two-dimensional inverse problem formulated under the assumptions of the Born approximation [18] within the scattering region. Sources and receivers will be located on an identical closed loop boundary surrounding a scattering region. Once selected, Green's functions and their normal derivatives along the measurement curve are used to project the field into the exterior region. A line of virtual sources is first formed through successive projections. Then, by argument of reciprocity, a line of virtual receivers parallel to the sources is formed, putting the data in a form suitable for well-established diffraction tomography methods [18].

In the reconstruction performed here, some variation in traditional approaches is taken in order to better exploit information available from having sources/receivers that entirely surround the object. Since the angle of orientation of the virtual source/receiver planes is arbitrary, multiple sets of virtual arrays may be formed from a single acquisition, allowing construction of spatial frequency components out to a radius equal to twice that of the imaging wavenumber. A k-space mapping scheme is introduced to minimize redundancy while maximizing the available information during this construction. Foundations of the approach are outlined, followed by the presentation of numeric examples.

## 2. Theory

### 2.1. Tomographic method

To begin, a continuously-radiating monochromatic point source at position $\mathbf{r}_S$ is considered. The radiated field is scattered by a spatially-varying but localized inhomogeneous region $q(\mathbf{r}')$ under the assumption that the Born approximation holds everywhere within this region. This field is measured by a point receiver located at $\mathbf{r}_R$. Placing both $\mathbf{r}_S$ and $\mathbf{r}_R$ outside the scattering region, the acoustic pressure, $p$, at the receiver can be represented by,

$$p(\mathbf{r}_R;\mathbf{r}_S) = \int q(\mathbf{r}')g_n(\mathbf{r}'|\mathbf{r}_S)g_n(\mathbf{r}_R|\mathbf{r}')d^n\mathbf{r}', \tag{1}$$

where the source amplitude is normalized and n is the dimensionality of the problem [11]. By convention, the vector following the semicolon in $p$ represents the source location or scattering origin. Functions $g_n$ are the associated Green's functions of the Helmholtz equation,

$$\nabla_n^2 g_n + k_0^2 g_n = \delta(\mathbf{r} - \mathbf{r}'). \tag{2}$$

It will be desirable to write the Green's functions in an integral form, which can be achieved by first taking the Fourier transforms of both sides of (2), producing an equation of the form

$$\tilde{G}(\mathbf{k}';\mathbf{r}') = \frac{e^{-i\mathbf{k}'\cdot\mathbf{r}'}}{k_0^2 - |\mathbf{k}'|^2}. \tag{3}$$

Working in Cartesian coordinates, the inverse Fourier transform of (3) with respect to $k_z'$ gives

$$G(k_x',k_y',z)_{z'} = \frac{i}{2}\frac{e^{-ik_x'x'}e^{-ik_y'y'}e^{ik_z|z-z'|}}{k_z}; \ k_z = \sqrt{k_0^2 - k_x'^2 - k_y'^2}, \tag{4}$$

and though the argument could continue in three dimensions, most practical imaging configurations concern two-dimensions, so that here $k_y'$ will be set to zero. The inverse transform with respect to $k_x'$ can then be expressed as

$$g(\mathbf{r}|\mathbf{r}') = \frac{i}{4\pi}\int\frac{e^{ik_z|z-z'|}}{k_z}e^{ik_x'(x-x')}dk_x'. \tag{5}$$

To proceed, the scattering region is confined to $0 < z < z_R$, and $\mathbf{r}_S$ is restricted to $z_S \leq 0$, as shown in Figure 1. Making use of (5) to rewrite the Green's functions in (1), the equation becomes

$$p(\mathbf{r}_R;\mathbf{r}_S) = \int q(\mathbf{r}')\left[\frac{e^{-ik_{S_z}x_S}}{k_{S_z}}e^{ik_{S_z}(z'-z_s)}e^{ik_{S_x}x'}\right]\left[\frac{e^{-ik_{R_z}x'}}{k_{R_z}}e^{ik_{R_z}(z_R-z')}e^{ik_{R_x}x_R}\right]dk_{R_x}dk_{S_x}d^2\mathbf{r}'. \tag{6}$$

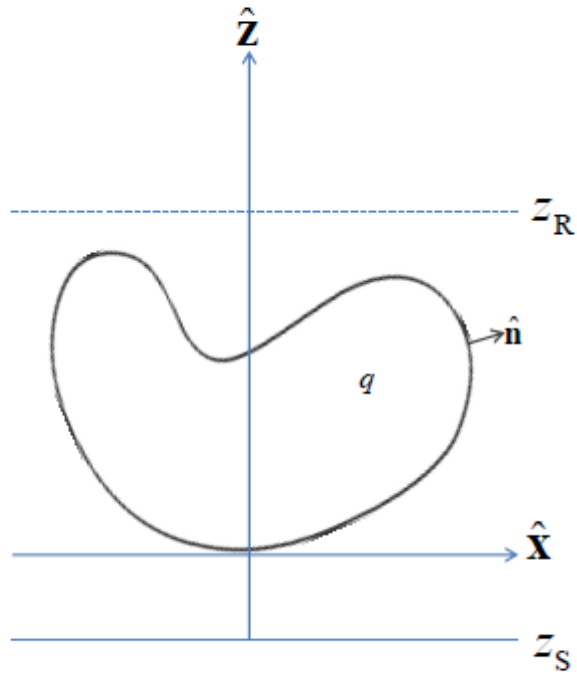

Figure 1

Configuration of the problem showing scattering region q located between a line of sources along $Z = Z_S$ and receivers $Z = Z_R$.

Rearranging terms and changing the order of integration puts (6) in the form of a two dimensional Fourier transform with respect to $\mathbf{r}'$,

$$p(\mathbf{r}_R;\mathbf{r}_S) = \int \frac{e^{-ik_{S_x}x_S}e^{-ik_{S_z}z_S}e^{ik_{R_x}x_R}e^{ik_{R_z}z_R}}{k_{S_z}k_{R_z}}\left[\int q(\mathbf{r}')\,e^{-i(k_{R_z}-k_{S_z})z'}e^{-i(k_{R_x}-k_{S_x})x'}\,d^2\mathbf{r}'\right]dk_{R_x}\,dk_{S_x}\ . \tag{7}$$

Solving the integral within the bracket yields

$$p(\mathbf{r}_R;\mathbf{r}_S) = \int \frac{Q(\mathbf{k}_R - \mathbf{k}_S)}{\mathbf{k}_{S_z}\mathbf{k}_{R_z}}\,e^{i\mathbf{k}_S\cdot(-\mathbf{r}_S)}\,e^{i\mathbf{k}_R\cdot\mathbf{r}_R}\,dk_{S_x}\,dk_{R_x}\ , \tag{8}$$

which takes on the form of an inverse Fourier integral. Thus, if $p$ is known along some line $x_S$ at an arbitrary constant $z_S$ ($z_S \leq 0$) for every point $x_R$ along an arbitrary constant $z_R$ ($z_R \geq 0$), the Fourier transform of the resulting function, $p(x_R;x_S)|_{z_R,z_S}$ can be equated with the integrand of the right hand side. Synthetic aperture tomographic approaches are based on direct measurement of

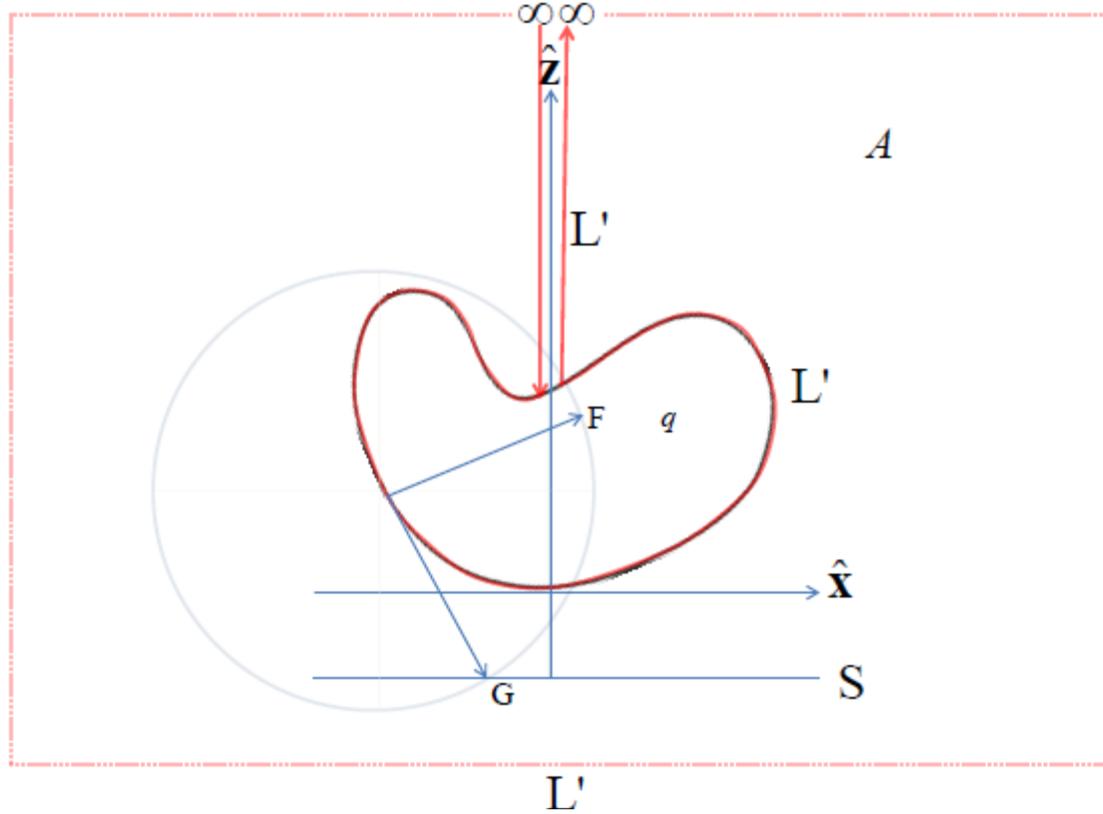

Measurement in the scattering region $q$ acquired with both sources and receivers located on L'. An interior Green's function, $F$, is shown at an appropriate point, given exterior Green's function $G$. If the diameter of $q$ becomes small relative to $G$, the method can be replaced by a series or numeric method. The measured signal is projected to S, producing a virtual array.

$p(x_R; x_S)$ to perform this inversion. Accessibility, however, may make it impossible to acquire data over such lines. Moreover, impractically-long lengths of sources $x_S$ and receivers $x_R$ may be required in order to achieve an accurate reconstruction. To overcome such restrictions, a more general case is now considered where signals may be acquired over any arbitrary, continuous, closed surface surrounding the inhomogeneous region.

### 2.2. Projection to separable boundaries

The integral theorem of Helmholtz and Kirchhoff in two dimensions [10] guarantees that for any point $\mathbf{r}_R$ located within the source-free homogeneous area $A$ enclosed by a curve L,

$$p(\mathbf{r}_R) = \frac{1}{2\pi} \oint_L \big( g(\mathbf{r}_R \mid \mathbf{r}')\nabla p(\mathbf{r}') - p(\mathbf{r}')\nabla g(\mathbf{r}_R \mid \mathbf{r}') \big) \cdot \hat{\mathbf{n}}' \, dl' , \qquad (9)$$

where the normal vector $\hat{\mathbf{n}}'$ is taken outward from the surface of integration and $p(\mathbf{r}')$ is the *total* field pressure on the curve. Selecting a curve that partially borders the scattering region, as

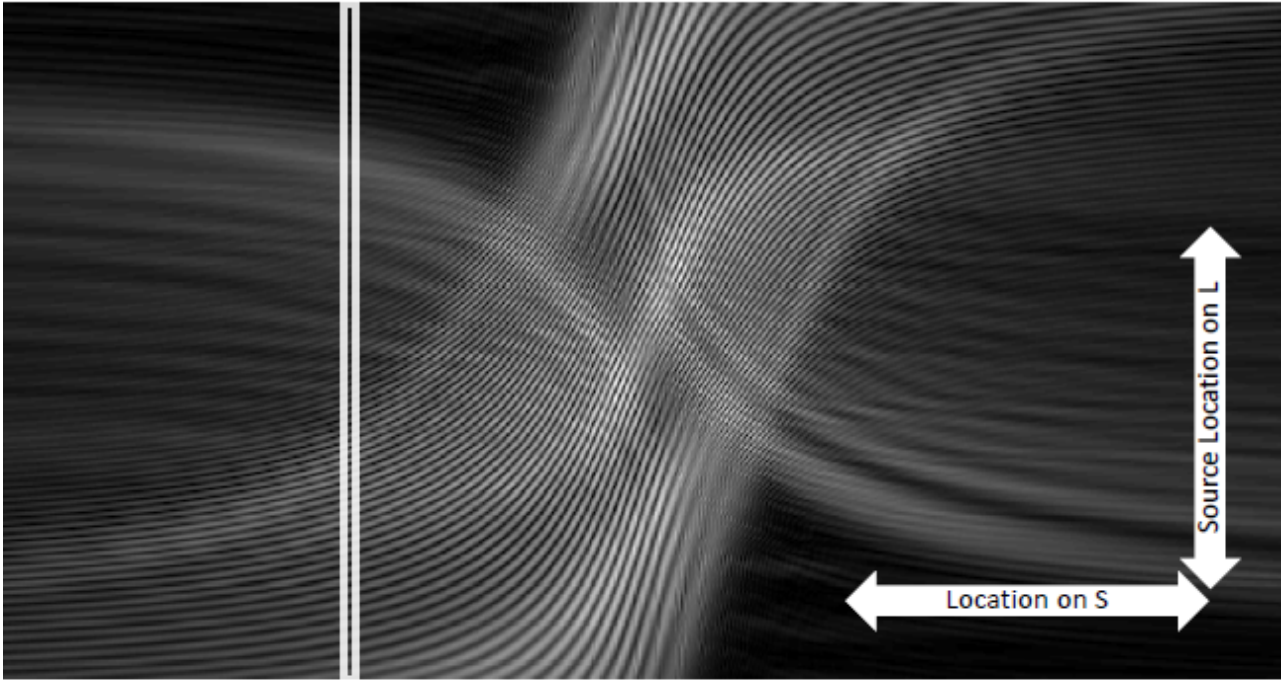

Figure 3

Modulus of the projected pressure along S as a function of source location on L. Values recorded at a fixed location on S (boxed) are equated to the pressure experienced on L from a hypothetical source at this location.

illustrated in Figure 2, shows the possibility to solve for $p(\mathbf{r}_R)$ over a finite line inscribed within the area bounded by L. If the medium is homogeneous outside the scattering region, the boundary away from the scattering region may be extended to infinity, reducing the integration bounds to L′, a section of the curve L surrounding the scattering region. This "medium" need not be a physical one, as long as measurements on L can be acquired without interference from the exterior region. For an incident field created by a point source located at $\mathbf{r}$, the resulting pressure at $\mathbf{r}_R$ gives:

$$p(\mathbf{r}_R;\mathbf{r}) = \frac{1}{2\pi} \int_{L'} \left( g(\mathbf{r}_R \mid \mathbf{r}') \frac{\partial p(\mathbf{r}';\mathbf{r})}{\partial n'} - p(\mathbf{r}';\mathbf{r}) \frac{\partial g(\mathbf{r}_R \mid \mathbf{r}')}{dn'} \right) \cdot \hat{\mathbf{n}}' dl' ,\qquad (10)$$

where $\partial / \partial n'$ is the normal derivative orthogonal to L′. A direct solution to this over-specified equation may be obtained, provided both the pressure and its normal derivative are known on the boundary and that care is also taken to avoid problems regarding non-uniqueness of the solution [17,19,12]. More typically pressure and its derivative are not simultaneously known and numeric boundary methods [14] must be used to find the unknown quantity.

The present goal is to maintain validity, even for measurement around complex boundary shapes. Here a unique solution to (10) is sought by applying Dirichlet conditions, such that $g(\mathbf{r}_R \mid \mathbf{r}')$=0 for $\mathbf{r}'$ on L. This requires an appropriate choice of Green's function, which is selected to be of the form

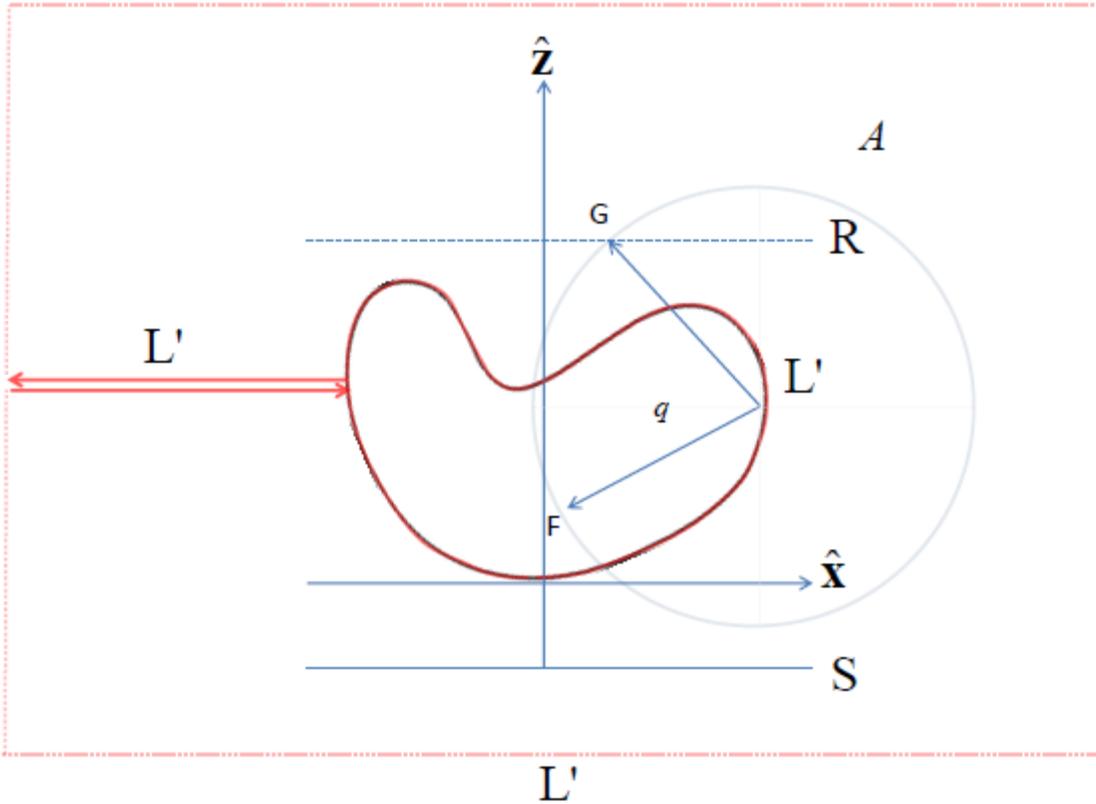

Figure 4

The virtual array at S is then used in combination with data on L′ to produce a virtual array of receivers along line R.

$$g(\mathbf{r_R}, \mathbf{r}) = g_0(\mathbf{r_R} \mid \mathbf{r}) + f(\mathbf{r_R}, \mathbf{r}),$$ (11)

where the first term is the outgoing Green's function of the Helmholtz equation. Clearly, in order for boundary conditions to be met $f(\mathbf{r_R}, \mathbf{r}') = -g_0(\mathbf{r_R} \mid \mathbf{r}')$ on L. Noting that the function $f$ must itself be a solution to the homogeneous Helmholtz equation, the function must also satisfy Green's second theorem [20], allowing (11) to be written

$$g(\mathbf{r_R}, \mathbf{r}) = g_0(\mathbf{r_R} \mid \mathbf{r})$$
$$+ \frac{1}{2\pi} \iint_{L'} \left( g_0(\mathbf{r} \mid \mathbf{r}') \frac{\partial f(\mathbf{r_R}, \mathbf{r}')}{\partial n'} - f(\mathbf{r_R}, \mathbf{r}') \frac{\partial g_0(\mathbf{r} \mid \mathbf{r}')}{\partial n'} \right) \cdot \hat{\mathbf{n}}' dl' .$$ (12)

Substituting $f = g - g_0$ into the integrand of (12) and allowing $\mathbf{r}$ to approach a point on the boundary, $g(\mathbf{r_R}, \mathbf{r} \to \mathbf{r}_0') \simeq 0$ yields a Fredholm integral of the first-kind

$$g_0(\mathbf{r}_R \mid \mathbf{r}_0') + \frac{1}{2\pi} \int_{L'} g_0(\mathbf{r}_0' \mid \mathbf{r}') \left\{ \frac{\partial g(\mathbf{r}_R, \mathbf{r}')}{\partial n'} \cdot \hat{\mathbf{n}}' \right\} dl' \Big|_{\mathbf{r} \to \mathbf{r}_0'} = 0, \tag{13}$$

amenable to inversion [21] in order to provide the unknown normal derivative at position $\mathbf{r}_0'$. Assuming, for now, that this normal derivative can indeed be determined by inversion (or otherwise approximated), (10) simplifies to

$$p(\mathbf{r}_R; \mathbf{r}) = \frac{1}{4\pi} \int_{L'} p(\mathbf{r}'; \mathbf{r}) K(\mathbf{r}_R, \mathbf{r}') dl', \tag{14}$$

where on the boundary

$$K(\mathbf{r}_R, \mathbf{r}') = \left\{ \frac{\partial g(\mathbf{r}_R, \mathbf{r}')}{\partial n'} \cdot \hat{\mathbf{n}}' \right\}. \tag{15}$$

Since (15) is dependent only on $\mathbf{r}_R$ and the boundary geometry, it is notably independent of source location. Thus once determined for a given external location, the same function is applicable to any source point, or array of points, on the boundary.

A solution along the line of constant $z_R$ can now be obtained for a continuum of points located on S.

By reciprocity, a point source radiating *from* a given location o n S would result in a pressure $p(\mathbf{r}; \mathbf{r}_R)$ on L (Figure 3). Renaming $\mathbf{r}_R \to \mathbf{r}_S$,

$$p(\mathbf{r}; x_S)|_{z_s} = \frac{1}{4\pi} \int_{L'} p(\mathbf{r}'; \mathbf{r}) K(\mathbf{r}_S, \mathbf{r}') dl', \tag{16}$$

giving the field on L, were a source to exist at a point on S. By confining S to the region $z_S \leq 0$ and applying (13) once again for a closed curve that fully circumscribes a line R parallel to S confined to $z_R \geq z_{q_{max}}$ yields

$$p(x_R; x_S)|_{z_R, z_S} = \frac{1}{4\pi} \int_{L''} p(\mathbf{r}''; \mathbf{r}_S) K(\mathbf{r}_S, \mathbf{r}'') \cdot \hat{\mathbf{n}}' dl''. \tag{17}$$

Finally, combining (16) and (17) into a single equation gives

$$p(x_R; x_S)|_{z_R, z_S} = \frac{1}{16\pi} \int_{L', L''} p(\mathbf{r}'; \mathbf{r}'') K(\mathbf{r}_R, \mathbf{r}'') K(\mathbf{r}_S, \mathbf{r}') dl' dl'', \tag{18}$$

as illustrated in Figure 4.

By (18) it is seen that an array of point sources and receivers assembled around an identical closed curve surrounding a scattering region can be used to measure the pressure at points around the curve due to each source, thereby forming the solution $p(\mathbf{r}'; \mathbf{r}'')$ over the curve points $\mathbf{r}'$ and $\mathbf{r}''$. This solution is used to produce the virtual solution $p(\mathbf{r}_R; \mathbf{r}_S)$ that would exist were sources and receivers to be located on R and S, respectively (Figure 4), with lines on L arbitrarily extending to and from infinity normal to the z-axis, thus avoid intersection with $z_S$ and $z_R$.

It is again noted that the space comprising R and S need not be a physical space, but rather may be regarded as a mapping of $p$ for the purpose of placing the data in the form of the solution of (8), making them suitable for reconstruction. It is further noted that the choice of mapping to a linear boundary could, in essence, be replaced by a mapping to a circular geometry to be reconstructed [22–25]. The current choice of Cartesian symmetry is based on the reconstruction method described in the next section.

### 2.3. Reconstruction

Equating the Fourier transform of (18) with respect to $x_R$ and $x_S$ to the integrand of (8), leads to

$$Q(\mathbf{k}_R - \mathbf{k}_S) = P(k_{R_x}, k_{S_x}) k_{S_z} k_{R_z} e^{i k_{S_z} z_S} e^{-i k_{R_z} z_R} . \tag{19}$$

Reconstruction entails the nonlinear mapping from the plane $(k_{R_x}, k_{S_x})$ to the Cartesian plane $(k_x, k_z)$ over which the object's spatial Fourier transform is specified. This mapping requires careful consideration of the resolution and range required in each space in order to produce an image of a given resolution. For clarity, the coordinates of the transform of $P$ are denoted as the "$K_{RS}$" space $(k_{R_x}, k_{S_x})$, whereas the Fourier transform of the object, $q$, is indicated as the "$K_q$" space $(k_x, k_z)$. A number of works have considered the optimal means to perform this mapping [26,27], which is described by

$$\begin{aligned} k_x &= k_{R_x} - k_{S_x} \\ k_z &= \sqrt{k_0^2 - k_{R_x}^2} - \sqrt{k_0^2 - k_{S_x}^2} \end{aligned}, \tag{20}$$

and whose key features are illustrated in Figure 5. It is noted that $k_{R_x} = k_{S_x}$ maps to the origin, while $k_{R_x} = -k_{S_x}$ maps to $k_z = 0$ over the span $-2 k_0$ to $2 k_0$, indicating the ability to conceivably reconstruct $K_q$ over a diameter of twice that of $K_{RS}$.

While algorithms typically map points directly from $K_{RS}$ to $K_q$, followed by interpolation to a linear grid, the present algorithm elects to algebraically solve for $k_{R_x}$ and $k_{S_x}$ in terms of $k_x$ and $k_z$, which has solutions

$$\begin{aligned} k_{R_x} &= \frac{k_x}{2} \pm \frac{1}{2} \sqrt{\frac{4 k_0^2 k_z^2 - k_x^2 k_z^2 - k_z^4}{k_x^2 + k_z^2}} \\ k_{S_x} &= k_{R_x} - k_x \end{aligned}. \tag{21}$$

Both solutions (±) of $k_{R_x}$ are relevant, with the physical solution dependent upon the particular location in $K_q$. When $k_{R_x} > k_x$ ($k_{S_x} > 0$) the additive solution (+) is used for Quadrants II and IV, the subtractive solution (-) for quadrant III, and no solution exists for quadrant IV. When $k_{R_x} < k_x$ ($k_{S_x} < 0$) the subtractive solution is used in quadrants I and III, the additive solution in quadrant IV, and no solution exists in quadrant II. By pre-specifying the area and spatial resolution of the image to be constructed, The desired points over a regular grid in $K_q$ can be mapped using (21) to provide the desired points in $K_{RS}$. Once specified, the value Q for a given point can be determined using (19). Finally, Q can be inverse Fourier transformed to produce an image. Conversely, determining Q from (20) maps values from $K_{RS}$ onto a nonlinear array of points in $K_q$ that is highly oversampled near the origin. This requires interpolation to a regularly spaced grid, which can both decrease processing efficiency and reduce accuracy [27,28]. An example comparing (20) and (21) is provided in Sec. 3.3.

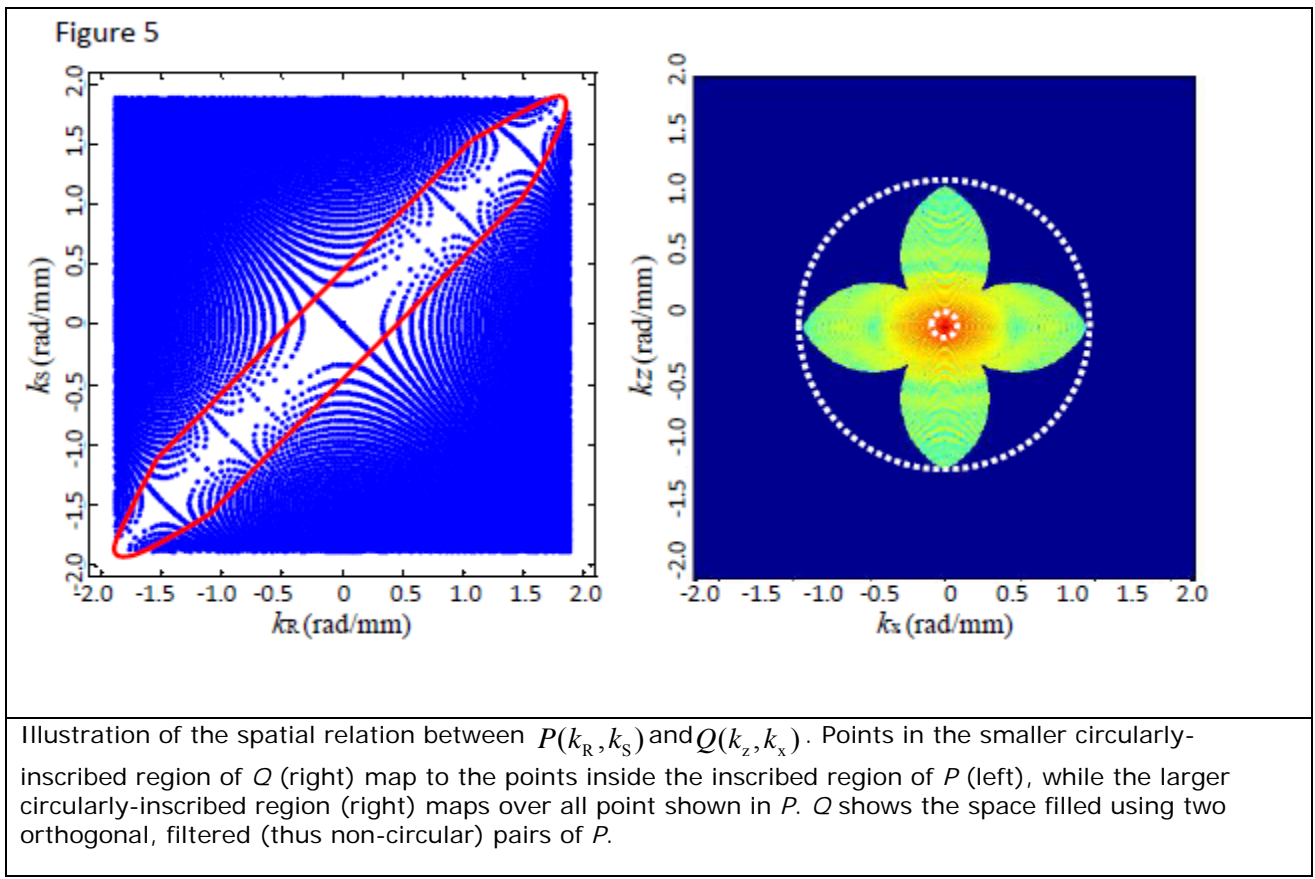

Figure 5

Illustration of the spatial relation between $P(k_R, k_S)$ and $Q(k_z, k_x)$. Points in the smaller circularly-inscribed region of $Q$ (right) map to the points inside the inscribed region of $P$ (left), while the larger circularly-inscribed region (right) maps over all point shown in $P$. $Q$ shows the space filled using two orthogonal, filtered (thus non-circular) pairs of $P$.

## 3. Numerical examples

### 3.1 Algorithm

To demonstrate the imaging process, a discrete approximation to the methods of Sec. 2 is presented. Calculations are implemented in Matlab (R2010a, Mathworks, Inc.) starting from a measured acoustic signal over an array of point-like sources/receivers on a closed curve that surrounds an ROI. The absolute positions of the points are assumed known. Vectors normal to the curve are determined by taking the cross product between tangential vectors (determined by differences between neighboring position vectors) and a vector normal to the imaging plane. One source is radiated and the resulting scattered field is recorded over all receivers. The process is repeated until all points have served as a source.

Green's functions are next determined to associate points on the curve to those along a virtual line source at $(x_s, z = 0)|_\phi$, where $\phi$ denotes a relative rotation angle. Gradients of the Green's functions normal to the measurement curve are similarly calculated and stored. In practice, these functions can be calculated *a priori* using (11) and stored in RAM for more efficient implementation.

With these values, signals are projected to a line along $z = S$ by a discrete approximation to (14). By repeating the projections over all source locations, the 2D matrix $p(x_s; r')|_{z_S=0}$ is formed. By acoustic reciprocity, a column of the transpose of this matrix can be equated to the field on the curved surface $r'$ that results from a virtual source at point $x_s$ as illustrated in Figure 3. This new field, $p(r'; x_s)$ can then be projected to a new line, $z = R$ using (16) and (17). This process is repeated over all $x_s$ to form the matrix $p(x_R; x_S)|_{z_S=0}^{z_R=Z}$.

Based on the size of the VOI, the desired spatial resolution, and the imaging frequency, the algorithm determines the necessary grid sizing and the number of virtual source/receiver rotation angles necessary to construct $Kq$. As there is significant redundancy between data from any two rotation angles, this procedure utilizes only $k_{R_x}$ and $k_{S_x}$ values that contribute uniquely to the uniform grid of $Kq$ described by (21). For example, a relatively simple and low resolution reconstruction might require only a single pair of virtual sources/receivers. If a single pair is deemed insufficient, missing values are next sought from a second virtual pair rotated $\pi/2$ relative to the first. This is followed, as necessary, by additional pairs rotated $\pm\pi/4$, then 4 additional pairs rotated $\pm\pi/8$ and $\pm 3\pi/8$, etc., until the preselected area of $Kq$ is obtained to within a prescribed resolution.

Since only specific and nonlinearly-spaced values of $k_{R_x}$ and $k_{S_x}$ are required by (21), it is neither necessary nor advantageous to calculate full Fourier transforms of $p(x_R; x_S)$. Rather, the values pertaining to a specific location $(k_x, k_z)$ are calculated directly in terms of $(x_R, x_{S_x})$ by discrete formation of (19) in terms of the Fourier integral:

$$Q(k_x(k_{R_x}, k_{S_x}), k_z(k_{R_x}, k_{S_x})) = \left[ \int_{-\infty}^{\infty} p(x_R, x_{S_x}) e^{-ik_{R_x} x_R} e^{-ik_{S_x} x_S} dx_S dx_R \right] k_{S_z} k_{R_z} e^{ik_{S_z} z_S} e^{-ik_{R_z} z_R} . \quad (22)$$

In other words, the integral is solved only for values of $(k_{R_x}, k_{S_x})$ pre-specified by (21), so that only the minimum necessary number of values are calculated, decreasing both the number of required calculations and the required memory. For each rotation, only non-overlapping regions are calculated, so that for each rotation the number of necessary values decreases.

As (22) is performed in lieu of the Fourier transform (FFT) - inarguably a remarkably efficient operation - the benefit of this process may at first be unclear. Indeed, a discrete FFT of can generally be calculated more rapidly than a discretized (22), even when a relatively small number of values $(k_{R_x}, k_{S_x})$ are required [29]. A space of $p(x_R; x_{S_x})$ discretized grid of $N_k = N_R \times N_S$ points could be transformed by FFT in a time proportional to $N_k$ operations. Conversely, a discretized Fourier integral would require the same number of operations just to calculate a single point $(k_{R_x}, k_{S_x})$. The substantial advantages of the proposed method, rather, come from avoiding the subsequent mapping and interpolation required of the forward approach. Selecting exclusively the $N_O$ discrete points that map to the grid $(k_x, k_z)$, the mappable space is complete after $\frac{\pi}{4} N_O \times N_k$ operations. Conversely, the same calculation determined from an FFT and forward mapping results in a highly oversampled space near the origin that then requires interpolation.

A numeric and visual comparison between the two approaches is given in Sec. 3.5. For this example, a common efficient interpolation approach making use of Delaunay triangulation is used. If well implemented, the triangulation might be performed in as few as $N_O \log(N_O)$ operations [30]. This step is then followed by interpolating the values of each point mapped from $N_k$. The total expected time would then be that of approximately $2 N_k + N_O \log(N_O) + 3 N_O N_k$ operations.

As high spatial frequency components of the signals in $K_{RS}$ tend to be both decreased in magnitude and increased in noise content, lowpass filtering is used. Qualitatively the effect of this filtering is to transform the circular regions in $Kq$ into lens-shaped geometries. In implementing (22) an adjustable Butterworth lowpass filter is set (here, to a default value of $0.9\ k_0$) to remove high-frequency noise. The filtered signal is finally inverse-transformed to produce an image.

### 3.2 Parameters
The parameters for setting spatial resolution and image size may be elucidated by examining $K_q$. Selecting some desired image resolution, $\delta r$ in both directions, Nyquist criteria sets a requisite maximum (spatial) frequency of $k_{NYQ} = \pi / \delta r$ for $k_x$ and $k_z$. Conversely, length, $L$, and height $H$, of the imaging region set the resolution in $K_q$ at $\Delta k_x \leq \pi / L, \Delta k_z \leq \pi / H$. In the idealized case where $Q(\mathbf{k})$ from (19) is fully known up to $k_{NYQ}$ at the required resolution in both directions, total image recovery with the selected spatial resolution is possible. However, this would also require that the values of all points mapped from $K_{RS}$ onto $K_q$ are known.

Examining (20), it can be seen that an area close-to and centered-about the origin in $K_q$ maps to points over the full range of $K_{RS}$ (Figure 5). Thus, even an image consisting of relatively low spatial frequencies still requires knowledge of the higher spatial components of $K_{RS}$. The equation further shows that both $k_{R_x}$ and $k_{S_x}$ must be less than or equal to $k_0$ for $k_z$ to be real, confining the regions of possibly-known $K_q$.

The resolution requirement in $K_{RS}$ can be determined by taking the partial differential of (21),

$$\Delta k_{R_x} = \frac{\partial k_{R_x}}{\partial k_x} \Delta k_x + \frac{\partial k_{R_x}}{\partial k_z} \Delta k_z$$
$$\Delta k_{S_x} = \frac{\partial k_{S_x}}{\partial k_x} \Delta k_x + \frac{\partial k_{S_x}}{\partial k_z} \Delta k_z \quad , \tag{23}$$

such that the constant resolution in $K_q$ is nonlinear in $K_{RS}$.

To complete the space $K_q$, one or more additional acquisitions can be performed along axes rotated relative to the initial acquisition. For this purpose, the present approach proves advantageous:

Since sources and receivers surround the target, the relative angle of the virtual source/receiver lines is arbitrary, and any number of projections may be performed at different rotation angles. Using a continuum of rotation angles would ultimately permit $K_q$ to be constructed within a circular region of radius 2 $k_0$, centered about the origin.

### 3.3 Scattering Phantoms

Numeric phantoms were used to illustrate the approach as well as to verify the algorithm. These phantoms consisted of a set of two-dimensional maps that provided speed of sound as a function of position for a scattering object situated in an otherwise uniform medium. The boundaries of two Shepp-Logan phantoms [31] were modified to provide significant local curvature (Figure 6). In particular, appreciable concave features were added to the boundaries, representing a known potentially-significant source of error for many exterior approaches [32].

Distance and resolution phantoms were also utilized, which contained point and line scatterers based on typical ultrasound resolution and distance test phantoms, e.g. [33]. Test patterns contained two rows and two columns of crossing point-like scatteres, located at regularly decreasing intervals (56mm, 28 mm, 14 mm, 7 mm, 3.5 mm, 1.75 mm, and 0.88 mm) along each row/column, and two solid crossing lines. Based on a signal frequency of 500 kHz and a background sound speed of 1500 m/s and scattering speed of 1550 m/s, three boundary shapes were considered, each selected to represent significant variation in curve shape over the scale of several wavelengths (Figure 7) and variable distance of the scattering locations relative to the measurement surface, while remaining marginally large enough to utilize the approach without the need for iterative representation of the Green's functions [17].

Simulations were limited to 2 virtual source/receiver pairs, covering the region in k-space shown in Figure 3. In this manner the case of three curves was considered representing successively increased asymmetry. Vectors normal to the phantom boundary were determined by calculating the cross product between a vector perpendicular to the imaging plane and the vectors tangential to the curve, which were determined by differences between neighboring position vectors.

### 3.4 Green's functions

Success of the method is dependent on the proper choice of the function $g(\mathbf{r}_\mathrm{R}, \mathbf{r}')$ and, more directly, an accurate solution of (15); its normal derivative on the boundary. Despite a wide range of boundary approaches, most methods require that either the boundary not deviate significantly from a separable boundary, i.e. that the solution may be regarded separable [34], or that impedance boundary conditions hold [11]. For the two dimensional cases considered here, the boundary conditions are satisfied by the class of functions

$$f(\mathbf{r}_\mathrm{R}, \mathbf{r}') = -g_0(\mathrm{k}\,\mathbf{R}); \ \mathbf{R} = \left|\mathbf{r}_\mathrm{R} - \mathbf{r}'\right|(\cos\theta\,\hat{\mathbf{x}} + \sin\theta\,\hat{\mathbf{z}}) \tag{24}$$

for any $\theta$ where $\mathbf{R} \notin A$. In the present two-dimensional examples, $g_0(\mathbf{r}_\mathrm{R} \mid \mathbf{r}') = H_0(\mathrm{k}\left|\mathbf{r}_\mathrm{R} - \mathbf{r}'\right|)$, the zero-order Hankel function of the first kind expressed as a function of wavenumber, k, and spatial separation. It is noted that the function

$$f(\mathbf{r}_\mathrm{R}, \mathbf{r}') = -g_0(2\mathbf{r}' - \mathbf{r}_\mathrm{R} \mid \mathbf{r}') \tag{25}$$

Figure 6

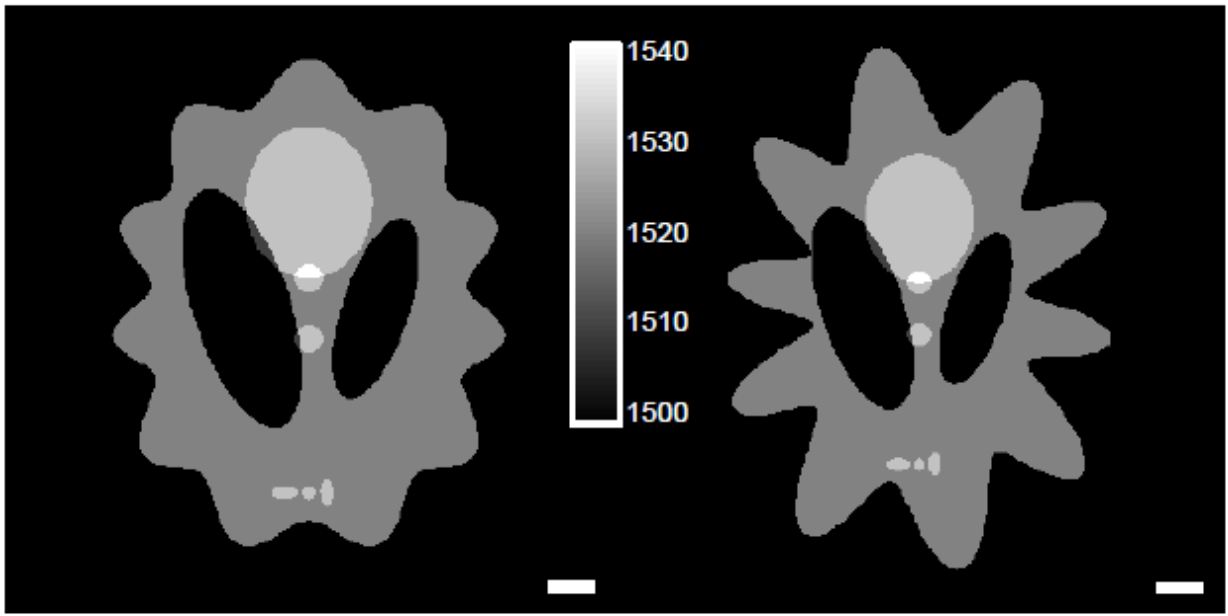

Shepp-Logan phantom varied in sound speed (m/s) with boundaries modified to represent two cases of significant local curvature. Horizontal bars indicate 10 wavelengths.

Figure 7

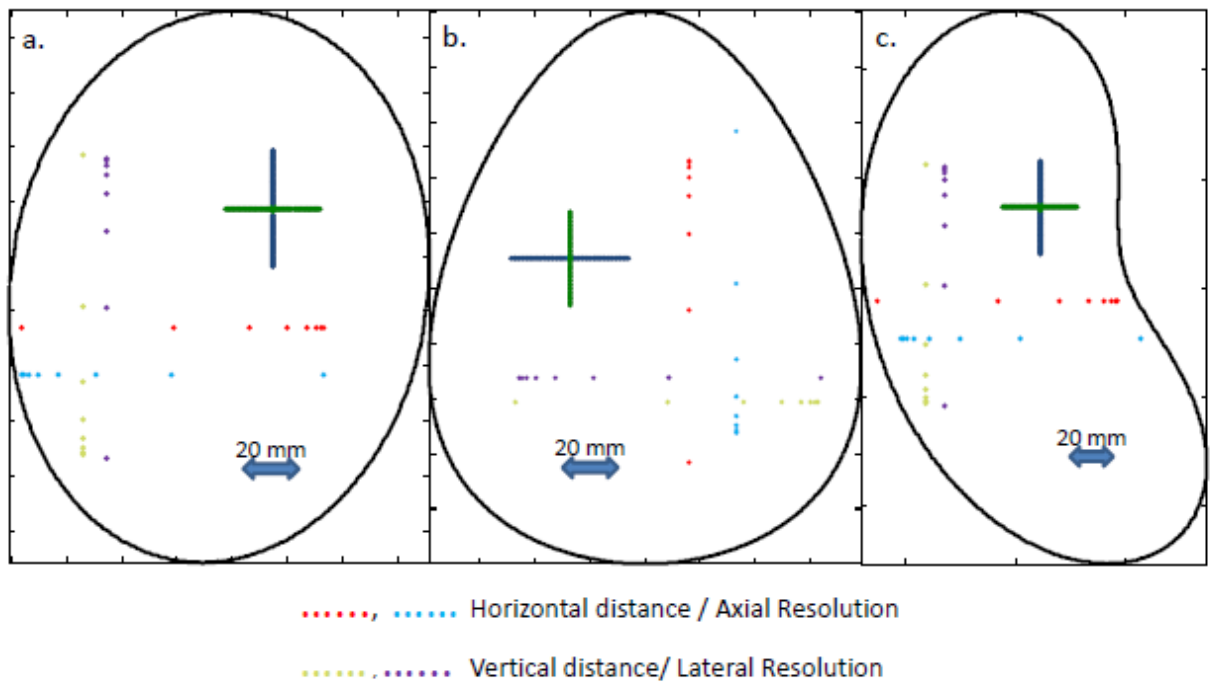

Horizontal distance / Axial Resolution

Vertical distance/ Lateral Resolution

Three curves defining the locations of sources and receivers used in the examples.

satisfies the necessary conditions for many practically-encountered curves, and is considered here for the distance and resolution phantoms described in Sec. 3.3. In this case, the solution to (15) on the boundary is formed by making use of the normal derivative,

$$\frac{\partial}{\partial n'} g_0(k|\mathbf{r}_R - \mathbf{r}'|) = \frac{\mathbf{r}_R - \mathbf{r}'}{|\mathbf{r}_R - \mathbf{r}'|} H_1(k|\mathbf{r}_R - \mathbf{r}'|) \tag{26}$$

with the derivative of $f$ calculated point-by-point over the boundary for a given $\mathbf{r}_R$.

When the local radius of curvature approaches the size of the imaging wavelength, as is the case for the modified Shepp-Logan Phantoms, a more general representation becomes necessary. This might be achieved by a number of boundary methods, and in the present case the normal derivative on the boundary is determined by applying the extinction theorem [19,35,36] in a manner identical to solving scattering of an external point source from a pressure release boundary. Making use of (15) and extending (12) to demand the exterior greens function be zero everywhere on the interior, the two-dimensional case,

$$\frac{1}{2\pi} \int_{L'} g_0(\mathbf{r}|\mathbf{r}') K(\mathbf{r}_R, \mathbf{r}') dl' = -H_0(k|\mathbf{r} - \mathbf{r}_R|), \qquad \mathbf{r} \in q. \tag{27}$$

A solution is obtained by establishing a system of integral equations formed by random selection of a set of locations within $q$. Solutions are then obtained using a generalized minimum residual method [37].

### 3.5 Simulations

For each simulated case, a field calculation was performed to simulate a radiating point source on the boundary of the phantom. The resulting pressure and its normal derivative were recorded at half-wavelength intervals along the exterior bounds of the object. This process was repeated with the source moved along the curve until a data grid of data was formed, so that the field on the boundary due to source radiation at any location was known. Simulations entailed discrete calculation of (1) using the two-dimensional Green's functions. Pressure gradients were then determined by applying the gradient operator to (1). Moving the operator inside the integral,

$$\nabla_R p(\mathbf{r}_R; \mathbf{r}_S) = \int q(\mathbf{r}') g(\mathbf{r}'|\mathbf{r}_S) \nabla_R g(\mathbf{r}_R|\mathbf{r}') d^2\mathbf{r}', \tag{28}$$

both $g(\mathbf{r}_R|\mathbf{r}')$ and $\nabla_R g(\mathbf{r}_R|\mathbf{r}')$ could be calculated *a priori* and stored for efficient implementation. These values were used as input for the algorithm described in Sec. 3.1.

In all simulations, virtual source and receiver lines were separated by 220 mm, each consisting of 900 elements spanning a distance of 1000 mm ($\Delta x = 1.11$ mm). The first curve (Figure 7a) was 560 mm in circumference, and formed the bounds of a nearly-elliptic oval with a long axis of 200 mm and a short axis of 150 mm. Source/receiver locations were located at 0.97 mm intervals along the curve. The second curve (Figure 7b) was egg-shaped, 560 mm in circumference, with a long axis of 200 mm and a short axis of 157 mm. Source/receiver locations were placed at 0.6 mm intervals around the curve. The third curve (Figure 7c) was 655 mm in circumference, spanning distance of 253 mm on the Cartesian z-axis and 158 mm on the Cartesian x-axis. Source/receivers were situated at 1.1 mm intervals on the curve.

**Figure 8**

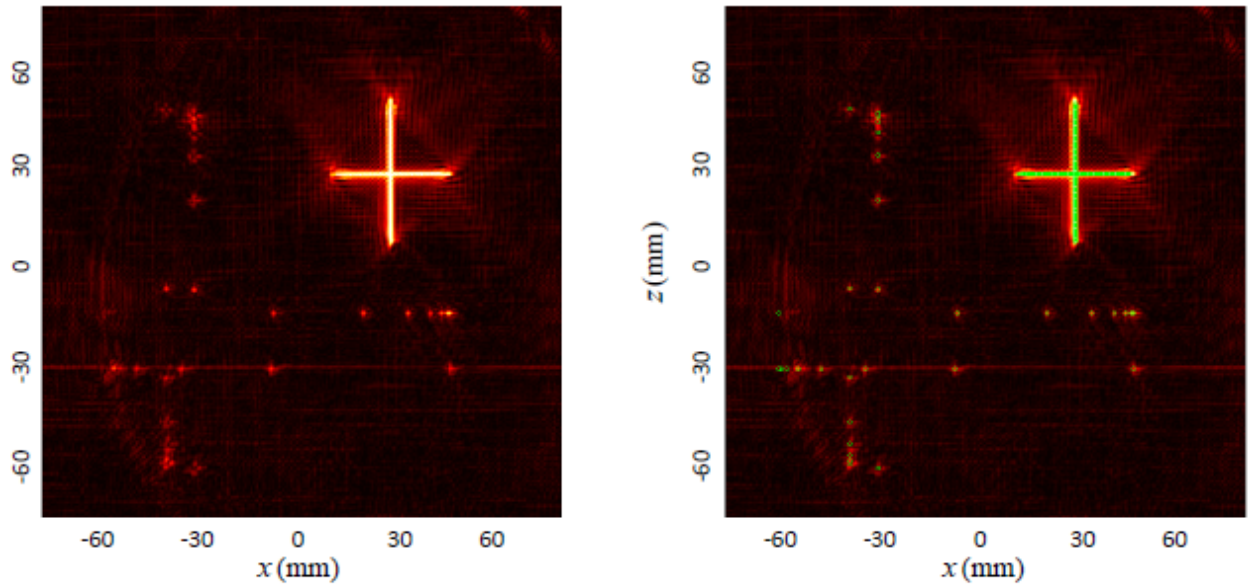

Image construction of the case shown in Figure 6a (left), and the same image overlain with the actual scattering locations (right).

**Figure 9**

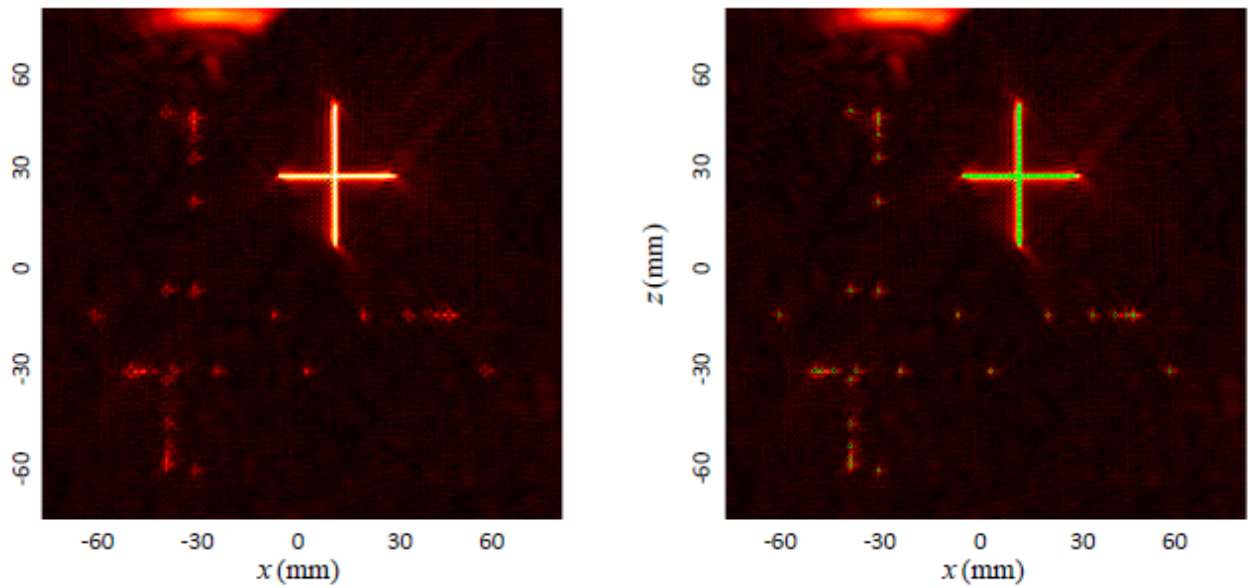

Figure 9. Image construction of the case shown in Figure 6b (left), and the same image overlain with the actual scattering locations (right).

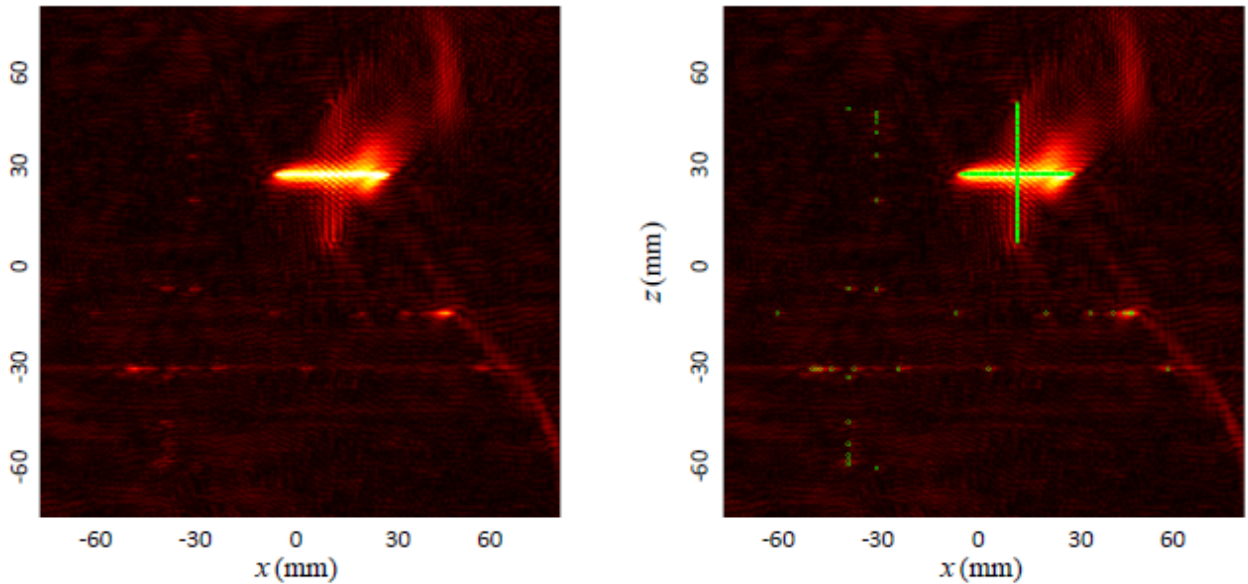

Figure 10. Image construction of the case shown in Figure 6c (left), and the same image overlain with the actual scattering locations (right).

Calculations were performed in Matlab R2010a using Windows 7 64-bit operating system. Hardware consisted of two quad-core E5-1620 3.6 GHz Xeon processors and 32 GB of RAM. Images were formed over a 512 X 512 grid spanning a 200 mm X 200 mm ROI. Results are displayed in Figures 8-10. In all three of these marginal examples, very little spatial distortion is observed over the span of the ROI, whereas the sub-wavelength scatterers are seen to exhibit point-spreading of one to several millimeters. Particularly in the third case (Figure 10), signal strength is diminished and significant blurring can be observed. However, even in this case most features of the phantom are discernible, the exceptions being when the point sources were very close to measurement plane. For the first phantom (Figure 7), spacing of 1.3 mm (~ 1/2 wavelength) or greater was discernible. For the second (Figure 8) and third phantom (Figure 9), spacing separation resolution was reduced, with spacing greater than 0.87 mm (~ 1 wavelength) discernible. Image artifacts are present in the latter two cases, as indicated in the figures.

Similar reconstructions of the modified Shepp-Logan Phantom are shown in Figures 11 and 12. Solutions in (27) were obtained in these cases using 4500 randomly-selected interior points from a uniform distribution. The Shepp-Logan phantoms were also utilized to examine relative efficiency of the algebraic approach expressed in (21), as compared to direct fast Fourier transformation (FFT) of the data into $K_{RS}$, mapping onto $K_q$ and then interpolation. The process time in implementing the FFT-based approach was observed to consistently benchmark at 11-14 times greater than the process time recorded using the algorithm described in Sec. 3.1. A qualitative comparison of the reconstructed images is also provided in the Figures.

**Figure 11**

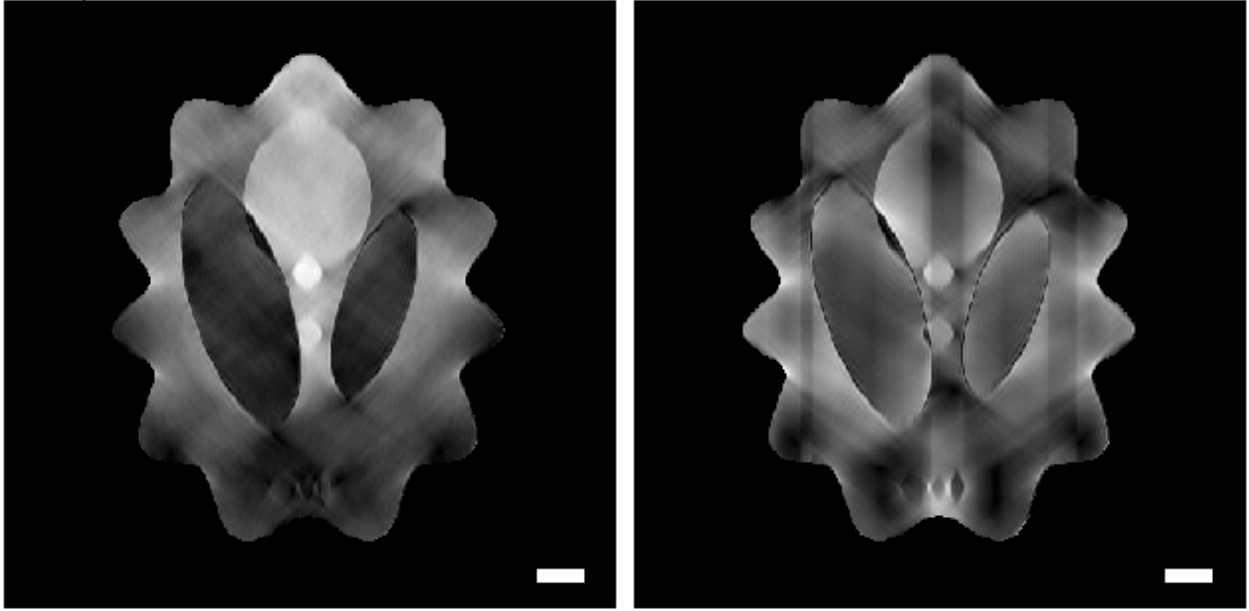

Image construction of the phantom in Figure 6(a). The left image was reconstructed using the mapping described in Sec. 3.1., while the right was formed by FFT, mapping, and interpolation. Bars indicate 10 wavelengths.

**Figure 12**

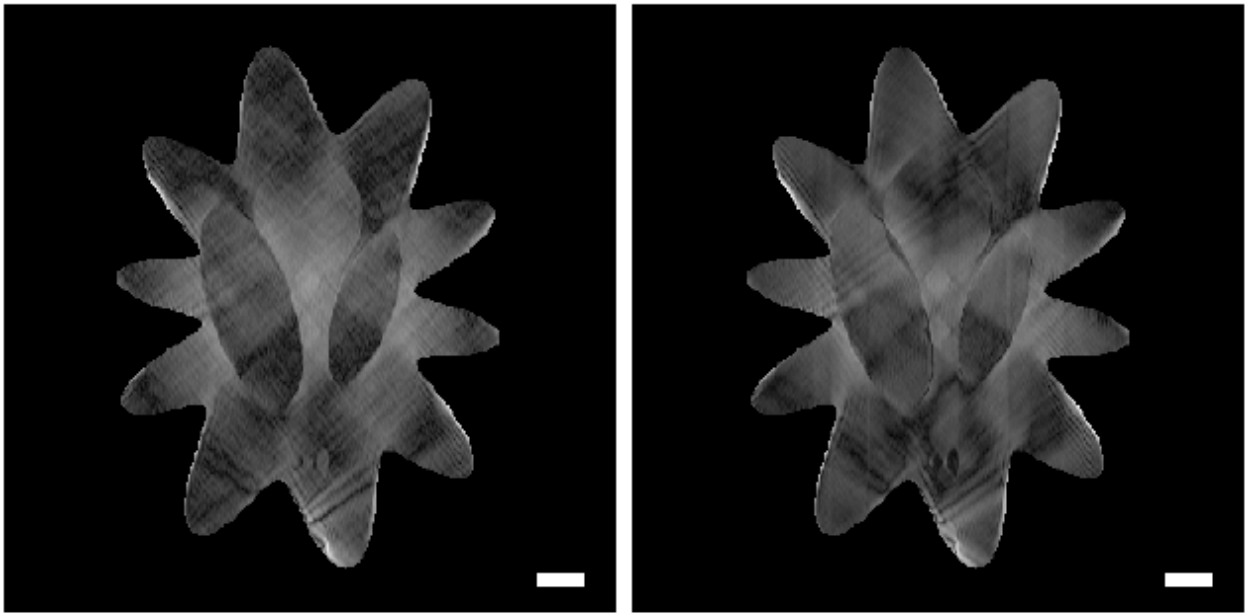

Image construction of the phantom in Figure 6(b). The left image was reconstructed using the mapping described in Sec. 3.1., while the right was formed by FFT, mapping, and interpolation. Bars indicate 10 wavelengths.

## 4. Summary

The tomographic process may be briefly summarized as follows: Starting from a closed curve of arbitrarily-shaped sources and receivers, a solution to the exterior problem under Dirichlet conditions is approximated, or otherwise numerically determined. Using this solution as the Green's function, received data is projected to a line exterior to the curve. This received and projected data is next associated, by reciprocity, to the signal expected on the original curve due to a virtual source on the external line. This new signal is further projected to a second line, denoted as the "receiver," placed parallel-to and on opposite sides of the object, thus reducing the problem to that of standard synthetic aperture diffraction tomography. Using only the original received dataset, this process may be repeated an arbitrary number of times at different angles.

This process was examined using two virtual source/receiver pairs formulated from three different irregular curves at 500kHz traveling through a medium comparable in sound speed to the range of many fluids or tissues. In this marginal case, point-scatterers could be identified, with very little distortion over the span of the reconstruction, albeit with low image contrast.

While this tomographic technique was specifically developed for application around irregularly-shaped boundaries, the method may have utility regardless of boundary shape in situations that would require sources and receivers to be placed over impractically-long linear distances in order to reconstruct an image. A mapping scheme was also described that was devised to optimize the construction in k-space by pre-selecting locations and associating their values with points from transformed data. While the method may have application in a number of areas in acoustics, our ongoing efforts are in applying a modified version of the approach toward tomographic ultrasound imaging of the brain.

## 5. Acknowledgements

Research reported in this publication was supported by the National Institute of Biomedical Imaging and Bioengineering of the National Institutes of Health under award number R01EB014296. The content is solely the responsibility of the author.

[1] Mueller R K 1980 Diffraction tomography I: The wave-equation *Ultrason Imaging* **2** 213–22

[2] Devaney A J 1982 A filtered backpropagation algorithm for diffraction tomography *Ultrason Imaging* **4** 336–50

[3] Greenleaf J F, Gisvold J J and Bahn R C 1982 Computed transmission ultrasound tomography *Med Prog Technol* **9** 165–70

[4] Natterer F and Wubbeling F 1995 A propagation-backpropagation method for ultrasound tomography *Inverse Problems* **11** 1225–32

[5] Harris J M 1987 Diffraction Tomography with Arrays of Discrete Sources and Receivers *IEEE Transactions on Geoscience and Remote Sensing* **GE-25** 448–55

[6] Mast T D 1999 Wideband quantitative ultrasonic imaging by time-domain diffraction tomography *The Journal of the Acoustical Society of America* **106** 3061–71

[7]  Mast T D, Feng Lin and Waag R C 1999 Time-domain ultrasound diffraction tomography *Ultrasonics Symposium, 1999. Proceedings. 1999 IEEE* Ultrasonics Symposium, 1999. Proceedings. 1999 IEEE vol 2 pp 1617–1620 vol.2

[8]  Devaney A J and Beylkin G 1984 Diffraction tomography using arbitrary transmitter and receiver surfaces *Ultrasonic Imaging* **6** 181–93

[9]  Gelius L-J 1991 A generalized diffraction tomography algorithm *The Journal of the Acoustical Society of America* **89** 523

[10]  Goodman J W 2005 *Introduction to Fourier optics* (United States: Roberts & Company)

[11]  Banaugh R P and Goldsmith W 1963 Diffraction of Steady Acoustic Waves by Surfaces of Arbitrary Shape *The Journal of the Acoustical Society of America* **35** 1590–601

[12]  Burton A J and Miller G F 1971 The Application of Integral Equation Methods to the Numerical Solution of Some Exterior Boundary-Value Problems *Proc. R. Soc. Lond. A* **323** 201–10

[13]  Meyer W L, Bell W A, Zinn B T and Stallybrass M P 1978 Boundary integral solutions of three dimensional acoustic radiation problems *Journal of Sound and Vibration* **59** 245–62

[14]  Banerjee P K, Banerjee P K and Butterfield R 1981 *Boundary element methods in engineering science* (McGraw-Hill Book Co. (UK))

[15]  Martin P A 1980 On the Null-Field Equations for the Exterior Problems of Acoustics *Q J Mechanics Appl Math* **33** 385–96

[16]  Brod K 1984 On the uniqueness of solution for all wavenumbers in acoustic radiation *The Journal of the Acoustical Society of America* **76** 1238–43

[17]  Ursell F 1973 On the exterior problems of acoustics *Mathematical Proceedings of the Cambridge Philosophical Society* **74** 117–25

[18]  Kak A C and Slaney M 2001 *Principles of Computerized Tomographic Imaging* (Society for Industrial Mathematics)

[19]  Schenck H A 1968 Improved Integral Formulation for Acoustic Radiation Problems *The Journal of the Acoustical Society of America* **44** 41–58

[20]  Sabatier E R P, Pierre C. 2001 *Scattering, Two-Volume Set: Scattering and Inverse Scattering in Pure and Applied Science* (Academic Press)

[21]  Polyanin A D and Manzhirov A V 2008 *Handbook of Integral Equations: Second Edition* (CRC Press)

[22]  Witten A, Tuggle J and Waag R C 1988 A practical approach to ultrasonic imaging using diffraction tomography *J. Acoust. Soc. Am* **83** 1645–52

[23]     Waag R C, Lin F, Varslot T K and Astheimer J P 2007 An Eigenfunction Method for Reconstruction of Large-Scale and High-Contrast Objects *IEEE Transactions on Ultrasonics, Ferroelectrics and Frequency Control* **54** 1316–32

[24]     Simonetti F and Huang L 2008 From beamforming to diffraction tomography *Journal of Applied Physics* **103** 103110–103110–7

[25]     Huthwaite P, Simonetti F and Duric N 2012 Combining time of flight and diffraction tomography for high resolution breast imaging: Initial invivo results (L) *The Journal of the Acoustical Society of America* **132** 1249–52

[26]     Pan S X and Kak A C 1983 A computational study of reconstruction algorithms for diffraction tomography: Interpolation versus filtered-backpropagation *IEEE Transactions on Acoustics, Speech and Signal Processing* **31** 1262–75

[27]     Bronstein M M, Bronstein A M, Zibulevsky M and Azhari H 2002 Reconstruction in diffraction ultrasound tomography using nonuniform FFT *IEEE Trans Med Imaging* **21** 1395–401

[28]     Devaney A J 1987 A fast filtered backpropagation algorithm for ultrasound tomography *IEEE Trans Ultrason Ferroelectr Freq Control* **34** 330–40

[29]     Vetterli M, Eurasip member and Nussbaumer H J 1984 Simple FFT and DCT algorithms with reduced number of operations *Signal Processing* **6** 267–78

[30]     Lee D T and Schachter B J 1980 Two algorithms for constructing a Delaunay triangulation *International Journal of Computer and Information Sciences* **9** 219–42

[31]     Shepp L A and Logan B F 1974 The Fourier reconstruction of a head section *IEEE Transactions on Nuclear Science* **21** 21–43

[32]     Bates R H T and Wall D J N 1977 Null Field Approach to Scalar Diffraction. II. Approximate Methods *Philosophical Transactions of the Royal Society of London. Series A, Mathematical and Physical Sciences* **287** 79–95

[33]     Pfeiffer D, Sutlief S, Feng W, Pierce H M and Kofler J 2008 AAPM Task Group 128: Quality assurance tests for prostate brachytherapy ultrasound systems *Medical Physics* **35** 5471–89

[34]     Chadan K, Colton D, Päivärinta L and Rundell W 1997 *An Introduction to Inverse Scattering and Inverse Spectral Problems* (SIAM)

[35]     Waterman P C 1969 New Formulation of Acoustic Scattering *The Journal of the Acoustical Society of America* **45** 1417–29

[36]     Bates R H T and Wall D J N 1977 Null Field Approach to Scalar Diffraction. I. General Method *Philosophical Transactions of the Royal Society of London. Series A, Mathematical and Physical Sciences* **287** 45–78

[37]     Saad Y and Schultz M H 1986 GMRES: A Generalized Minimal Residual Algorithm for Solving Nonsymmetric Linear Systems *SIAM J. Sci. Stat. Comput.* **7** 856–69